\documentclass[sigconf]{acmart}
\usepackage{listings}
\usepackage{subcaption}
\usepackage{balance}

\AtBeginDocument{%
  }

\setcopyright{acmlicensed}
\copyrightyear{2018}
\acmYear{2018}
\acmDOI{XXXXXXX.XXXXXXX}
\acmConference[FSE]{ACM International Conference on the Foundations of Software Engineering}{June 23--27, 2025}{Trondheim, Norway}

\begin{document}

\title{Capturing Semantic Flow of ML-based Systems}

\author{Shin Yoo}
\affiliation{%
  \institution{KAIST}
  \city{Daejeon}
  \country{Korea}
}
\email{shin.yoo@kaist.ac.kr}

\author{Robert Feldt}
\affiliation{%
  \institution{Chalmers University}
  \city{Gothenburg}
  \country{Sweden}}
\email{robert.feldt@chalmers.se}

\author{Somin Kim}
\affiliation{%
  \institution{KAIST}
  \city{Daejeon}
  \country{Korea}
}
\email{somin.kim@kaist.ac.kr}

\author{Naryeong Kim}
\affiliation{%
  \institution{KAIST}
  \city{Daejeon}
  \country{Korea}
}
\email{naryeong.kim@kaist.ac.kr}

\renewcommand{\shortauthors}{Yoo et al.}

\begin{abstract}
ML-based systems are software systems that incorporates machine learning components such as Deep Neural Networks (DNNs) or Large Language Models (LLMs). While such systems enable advanced features such as high performance computer vision, natural language processing, and code generation, their \emph{internal} behaviour remain largely opaque to traditional dynamic analysis such as testing: existing analysis typically concern only what is observable from the outside, such as input similarity or class label changes. We propose semantic flow, a concept designed to capture the internal behaviour of ML-based system and to provide a platform for traditional dynamic analysis techniques to be adapted to. Semantic flow combines the idea of control flow with internal states taken from executions of ML-based systems, such as activation values of a specific layer in a DNN, or embeddings of LLM responses at a specific inference step of LLM agents. The resulting representation, summarised as \emph{semantic flow graphs}, can capture internal decisions that are not explicitly represented in the traditional control flow of ML-based systems. 
We propose the idea of semantic flow, introduce two examples using a DNN and an LLM agent, and finally sketch its properties and how it can be used to adapt existing dynamic analysis techniques for use in ML-based software systems.
\end{abstract}

\received{20 February 2007}
\received[revised]{12 March 2009}
\received[accepted]{5 June 2009}

\maketitle

\section{Introduction}
\label{sec:introduction}

Recent rapid advances in machine learning, particularly in Deep Neural Networks (DNNs)~\cite{LeCun2015ef} and Large Language Models (LLMs)~\cite{Vaswani2017aa,Brown2020aa} have led to the emergence of ML-based software system, a new type of software systems that uses machine learning components such as DNNs or LLMs as part of the system. ML-based systems range from safety-critical systems that incorporate vision-related DNNs~\cite{Sharifi2023aa} to agentic systems driven by LLMs~\cite{Kang2024ay,Yang2024ab}. We expect such ML-based systems to be a prevalent form of software systems in the future, thanks to the unsurpassed capabilities of DNNs (in computer vision~\cite{Farabet2013bc,Feng2020mz} and speech recognition~\cite{Hinton2012aa}) and LLMs (in various natural language processing tasks~\cite{Brown2020aa}, text-based games~\cite{Gptrpg2024aa,Yan2023aa}, language education~\cite{duolingoMax}, as well as software engineering tasks~\cite{Fan2023yu}). 

With the increasingly wider adoption of ML-based systems, the need to analyse and verify their behaviour also increases, so that we can ensure the quality of service provided by these systems. However, it is not straightforward to apply existing program analysis techniques, such as testing and debugging, to ML-based systems, because the internal behaviour of the ML components (i.e., DNNs or Transformer models of LLMs) are fundamentally different from traditional software and therefore remain opaque. In machine learning literature, attempts to understand the internal behaviour of ML components tend to focus on identifying features of inputs~\cite{Zhang2018ab,Yun2021aa,Huben2024aa}. While these work help us understand ML components themselves better, they fall short of providing a perspective for a whole ML-based software system, which involves parts written as traditional software as well as ML components.

We propose a new representation of ML-system behaviour called semantic flow. Similarly to control and data flow for traditional software, semantic flow captures how semantic information in the latent space used by ML components changes as the system executes. It is different from existing attempt to capture input features, as semantic flow also captures the idea of system execution, traditionally represented in control flow. We present semantic flow using examples of both a DNN and an LLM-based agent system, show connections to existing work in testing of ML systems, and finally discuss future work on how traditional dynamic analysis techniques such as testing and fault localisation can be adapted to ML-based systems using semantic flow as a base representation.

\lstset{
  basicstyle=\footnotesize\ttfamily,
  columns=fullflexible,
  frame=lines,
  breaklines=true,
  postbreak=\mbox{\textcolor{gray}{$\hookrightarrow$}\space},
}
\begin{figure}[ht]
\begin{lstlisting}[language=Python]
class Net(nn.Module):
    def __init__(self):
        super().__init__()
        self.conv1 = nn.Conv2d(3, 6, 5)
        self.pool = nn.MaxPool2d(2, 2)
        self.conv2 = nn.Conv2d(6, 16, 5)
        self.fc1 = nn.Linear(16 * 5 * 5, 120)
        self.fc2 = nn.Linear(120, 84)
        self.fc3 = nn.Linear(84, 10)

    def forward(self, x):
        x = self.pool(F.relu(self.conv1(x)))
        x = self.pool(F.relu(self.conv2(x)))
        x = torch.flatten(x, 1)
        x = F.relu(self.fc1(x))
        x = F.relu(self.fc2(x))
        x = self.fc3(x)
        return x
\end{lstlisting}
\caption{A Convolutional Neural Network for Image Classification written in PyTorch\label{fig:cnn}}
\end{figure}

\section{Motivation}
\label{sec:motivation}

Let us first consider a simple image classification DNN, whose source code is listed in Figure~\ref{fig:cnn}. Both the model structure defined in \lstinline{__init__} and the actual computation defined in \lstinline{forward} are linear, i.e., lack any branching. Regardless of the classification results, all inputs follow the same execution path (i.e., the sequence of layers). However, classification expressed in traditional programming language would inherently based on branching, as shown in Figure~\ref{fig:classification}, because we use control flow to define program behaviour. In contrast, the CNN model in Figure~\ref{fig:cnn} performs the classification solely along the dataflow, i.e., by changing the distributions of activation values in the latent space.

\begin{figure}[ht]
\begin{lstlisting}[language=Python]
def classify(input):
    if features(input) == features["class_A"]:
      return "class_A"
    elif features(input) == features["class_B"]:
      return "class_B"
    elif features(input) == features["class_C"]:
      return "class_C"
    ...
\end{lstlisting}
\caption{A Classifier Logic expressed in Python\label{fig:classification}}
\end{figure}

Another motivating example we present is of AutoFL~\cite{Kang2024ay}, an agentic Fault Localisation (FL) technique that is driven by an LLM. At its core, AutoFL uses a ReAct~\cite{Yao2022qf} like function calling feature of GPT-3.5 and GPT-4 to overcome the limitations of context window lengths in LLMs. After being presented with the source code of the failing test case, the LLM is given autonomy over which function call to invoke. The LLM instance respond to the initial prompt with a series of requests to invoke one of the four available functions:

\begin{itemize}
\item \lstinline{get_class_covered}: returns the list of classes covered by the failing test case
\item \lstinline{get_method_covered}: given a class signature, returns the list of its methods covered by the failing test case
\item \lstinline{get_code_snippet}: given a method signature, returns the method body source code
\item \lstinline{get_comments}: given a method signature, returns the docstring that accompanies the method
\end{itemize}

\begin{figure}[ht]
\includegraphics[width=0.6\linewidth]{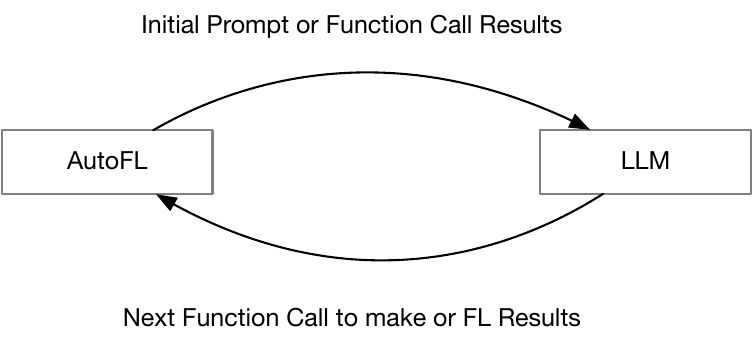}
\caption{Outline of AutoFL, an LLM-based FL agent~\cite{Kang2024ay}\label{fig:autofl}}
\end{figure}

Authors report that AutoFL can autonomously navigate the code repository, starting from the failing test case and gradually narrowing down to the location of the fault: it tends to start by looking at the list of covered classes, then the list of methods, and finally code snippets and comments of various methods, which is not unreasonable to human eyes. However, details of such behaviour can only be discerned in the semantic contents of LLM generated responses. From the perspective of traditional control flow, AutoFL simply repeats a loop between the source code that implements the function call invocations, and the LLM that makes such requests, as shown in Figure~\ref{fig:autofl}. The control flow alone does not reveal the rich semantic information, which is included in the contents of the function call requests made by the LLM. 

Based on these two examples, we argue that we need a new representation that combines the traditional control flow and the semantic information that drives ML components such as DNNs or LLMs. Such a representation would essentially serve as an alternative to the traditional concept of \emph{execution traces}, and allow us to investigate the behaviour of ML-based systems in a way that is more similar to traditional software. 

\section{Semantic Flow}
\label{sec:semantic_flow}


Semantic flow describes the sequential progression through one or more latent spaces within an ML component or ML-based system, analogous to how control flow describes the order of execution of program elements in traditional software. When given an input, an ML component (e.g., a deep neural network) or system (e.g., a large language model acting as an agent) traverses multiple latent spaces as part of its internal processing. These latent spaces represent the internal states and transformations that occur during execution.

We define a \emph{semantic state}, $s_i$, as a point in a latent space, encapsulating the representation of the system's internal state at a specific moment. A semantic flow is a sequence of such states, $s_1,s_2,\ldots,s_n$ where each $s_i$ resides in a latent space $LS_i$ corresponding to a distinct phase or step in the execution. For a DNN, $LS_i$ might represent the activations of a particular layer, while for an agentic system, $LS_i$ could denote the reasoning state at a specific step.

To generalise and analyse multiple executions, we aggregate semantic states into semantic clusters, forming a \emph{semantic flow graph} (SFG). In this graph, nodes represent clusters of semantically related states within one of the latent spaces, and edges capture transitions between these clusters observed---or theoretically possible---across a set of executions.

To construct semantic flows, three core elements are required:

\begin{itemize}
\item \emph{Unit of Analysis}: Define the specific execution steps to model and identify the relevant data structures for the internal state at each step. Let $e_i$ represent the execution data at step $i$.

\item \emph{Latent Mapping}: Specify a function
$embed(i, e_i)$ that maps execution data $e_i$ at step $i$ into a semantic state $s_i$ in the latent space $LS_i$.

\item \emph{Semantic aggregation}: Specify a semantic aggregation function, 
$aggregate(i, s_i, S_i)$, where $S_i$ represents all states mapped to $LS_i$. This function groups semantically related states into clusters and assigns each state to a node in the semantic flow graph. These clusters provide high-level semantic abstractions, improving interpretability and aiding analysis.
\end{itemize}

By formalizing semantic flows, we can better understand and visualize how ML systems progress through latent spaces during execution. This framework facilitates the comparison of system behaviour under different inputs or configurations, revealing similarities and differences in their logical and semantic decisions. We argue such insights can be useful for testing, optimization, and debugging, enabling targeted improvements in system behaviour. In the following, we present concrete examples of semantic flows and explore their practical applications.

\subsection{Semantic Flow of Image Classifiers}
\label{sec:image_classifier_flow}

\begin{figure}[ht]
\begin{subfigure}[t]{0.25\linewidth}
\includegraphics[width=25mm]{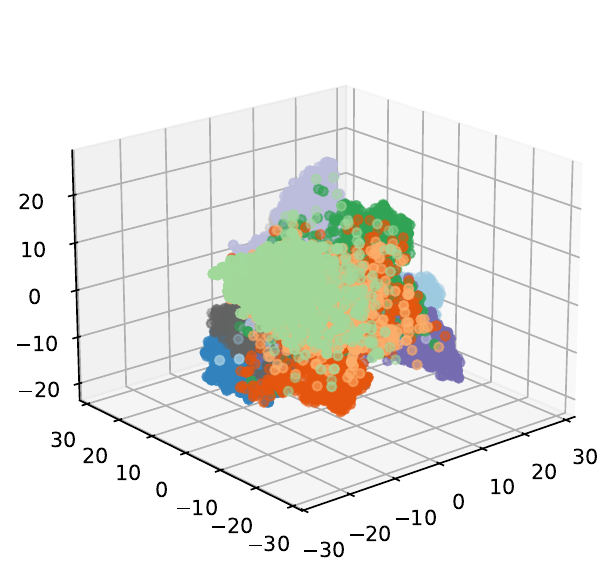}
\caption{FC Layer 1\label{fig:tsne1}}
\end{subfigure}
\begin{subfigure}[t]{0.25\linewidth}
\includegraphics[width=25mm]{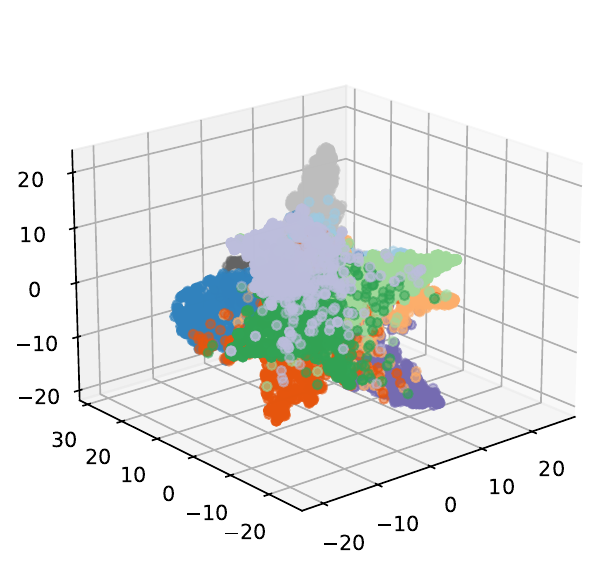}
\caption{FC Layer 2\label{fig:tsne2}}
\end{subfigure}
\begin{subfigure}[t]{0.25\linewidth}
\includegraphics[width=25mm]{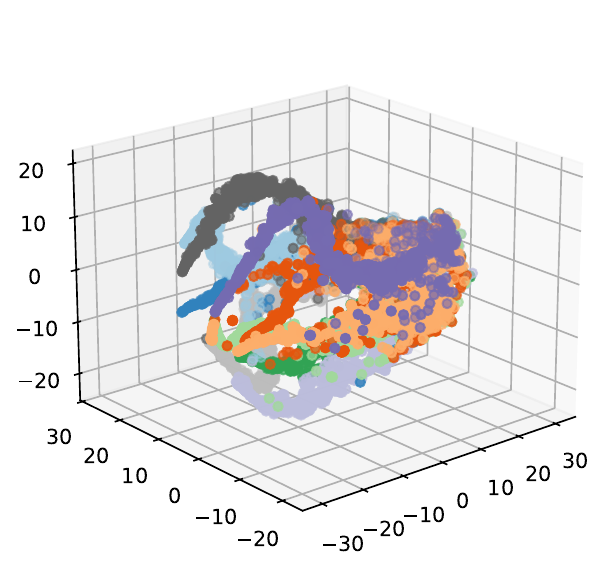}
\caption{FC Layer 3\label{fig:tsne3}}
\end{subfigure}
\caption{Semantic clusters observed at different layers of a CNN CIFAR-10 Classifier, visualised in 3D with t-SNE. Note that different classes are increasingly separated from each other as they pass layers, gradually forming more visually separated semantic clusters. We argue that such separations can be likened to decision branching in programs.\label{fig:tsne}}
\end{figure}

Consider the visualisation of semantic flow of a Convolutional Neural Network (CNN) trained to classify the CIFAR-10 benchmark~\cite{Krizhevsky2009kt}.
Figure~\ref{fig:tsne} visualizes the progression of semantic states that we extracted from the network's executions, with dots representing semantic states and their colour selected based on the network's predicted class such as birds, cats, ships, and aeroplanes.

The unit of analysis is the layers of the CNN and the semantic states are derived from the activation values in a layer during classification. For each input image, the network generates activations at successive layers, which reflect the system’s internal state at each step. We use the t-SNE algorithm to map these layer-wise activation values (execution data) into semantic states. Figures~\ref{fig:tsne1},~\ref{fig:tsne2}, and~\ref{fig:tsne3} show the semantic states from the three final fully connected layers of the network, each one reduced to (3D) latent spaces using t-SNE. As the input progresses through the layers, semantic states for each class become increasingly distinct, forming clusters by class label in the latent spaces. This separation can be seen to correspond to branching in the network's classification logic, as shown for traditional code in Figure~\ref{fig:classification}. Note that here we did not visualise the flow between the three latent spaces that would be needed to show the flow graph. Similarly, the aggregation function was trivial since it is only based on ground truth labels.

The clustering of semantic states we can observe aligns with findings by Rauber et al.~\cite{Rauber2017aa}, who observed that training improves class separation in latent space. Yosinski et al.~\cite{Yosinski2015oh} similarly demonstrated how visualizing activations can help reveal the role of convolutional layers. We build on these insights, and posit that the progression of internal activations through layers mirrors branching logic in traditional programs, where semantic flow captures how the network separates image classes for accurate classification.

\begin{figure}[h]
\includegraphics[width=65mm]{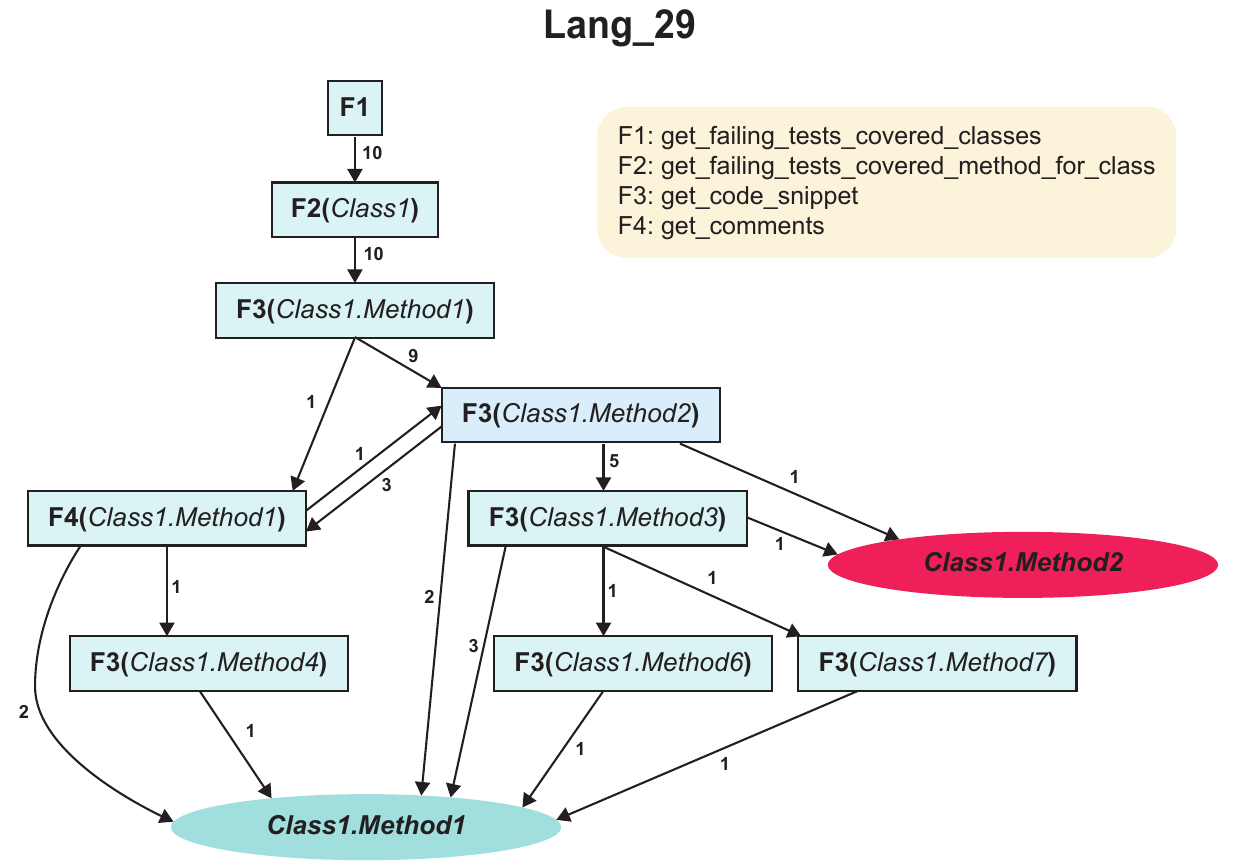}
\caption{LLM Inference Graph by Kim et al.~\cite{Kim2025aa} of FL inferences of AutoFL~\cite{Kang2024aa} for the bug Lang-29 of Defects4J~\cite{Just:2014aa}\label{fig:lig}}
\end{figure}

\subsection{Semantic Flow of LLM Agents}
\label{sec:llm_agents_flow}

LLM agents provide an intuitive context for semantic flow analysis. Their agentic workflows naturally define the ordering of information (akin to control flow), while their generated outputs encapsulate semantic information. The AutoFL system~\cite{Kang2024ay}, previously introduced, can serve as an illustrative case.

In AutoFL, semantic states are derived from the system's inference steps, where each step corresponds to a function call invoked by the LLM during fault localization. The system's execution sequence forms the basis for semantic flow analysis. Each function call is represented as a semantic state within a single latent space (shared between all steps), consistent with the LLM Inference Graph (LIG) framework by Kim et al.~\cite{Kim2025aa}. Semantic states are encoded discretely using one-hot representations, ensuring identical function calls with the same arguments map to the same state. Figure~\ref{fig:lig} visualizes an LIG derived from 10 AutoFL executions, showing how function calls progress through the workflow. The semantic states corresponding to identical function calls across executions are merged, with edge weights reflecting the frequency of transitions between them. For example, AutoFL's multiple executions for the same bug aggregate results for self-consistency~\cite{Wang2023aa}. In Figure~\ref{fig:lig}, some executions correctly localize the bug (reaching the blue node), while others fail (reaching the red node). These clusters and transitions clarify the logical pathways taken by the system.

The LIG depicted in Figure~\ref{fig:lig} highlights AutoFL's decision-making at each step, contrasting with traditional control flow (Figure~\ref{fig:autofl}). The LIG is a specific instance of a semantic flow graph, where semantic information is discretely encoded by function call types and the graph has been further compacted by counting and annotating with the number of node transitions. While in this case the semantic flow embedding function was manually selected we could alternatively use an external LLM encoder model to embed the natural language inputs\slash outputs into a latent space~\cite{Reimers2019aa} or directly using the hidden states of the LLM of the agentic system.
This analysis demonstrates how Semantic Flow Graph (SFG) analysis can provide a structured, interpretable view of LLM agent workflows.

\section{Properties and Applications of Semantic Flow}
\label{sec:applications}

Fundamentally, semantic flow aims to capture and represent the executions of ML-based systems at multiple latent spaces. Once we capture semantic flow of a set of accepted executions, it can be used to measure various properties of executions of ML-based systems. For example, we can imagine a diversity-aware testing technique for LLM-based agents that can select and prioritise inputs that result in the most diverse set of executions. We discuss potential applications of semantic flow in this section.

\subsection{Semantic and Control Flow Graphs (SaCFGs)} 

Semantic Flow Graphs (SFGs) represent execution flows as graphs, where nodes correspond to clusters of related states in a latent space, and edges indicate their sequential appearance. While these graphs can be helpful tools in themselves we also propose hybridizing SFGs with traditional Control Flow Graphs (CFGs) to create Semantic and Control Flow Graphs (SaCFGs) for analyzing hybrid systems that combine conventional software and machine learning (ML) components, such as large language models (LLMs). As one example, SaCFGs can enable defining and evaluating coverage criteria that span diverse components in complex systems.

Unlike static CFG analysis, constructing SaCFGs requires a multi-step process due to the stochastic nature of semantic state clustering, which depends on system executions. Once clusters are formed, new states can be mapped either as discrete assignments (e.g., $\epsilon$-coverage: states within distance $\epsilon$ of a cluster) or using finer-grained distances to measure partial coverage across multiple clusters. This hybrid framework offers a powerful tool for exploring semantic and control flow interactions in ML-software systems, advancing both analysis and practical applications.

\subsection{Measuring Out-of-Distribution-ness of Executions}
\label{sec:ood}


The statistical nature of semantic flows, particularly semantic clusters, provides a foundation for quantifying the out-of-distribution-ness of ML-based system executions. We argue that Surprise Adequacy (SA)~\cite{Kim2019aa,Kim2022hg}, a widely studied test adequacy metric for deep neural networks (DNNs), represents a specific instance of such a measure. SA evaluates how much of an outlier a new semantic cluster is relative to a reference cluster, with research showing that higher levels of surprise in inputs correlate with a greater likelihood of unexpected or buggy behavior in DNNs.

Extending this concept, semantic flow analysis enables out-of-distribution measurements for longer and more complex executions, such as those from systems involving large language model (LLM) agents. By considering multiple semantic states sampled at various points during execution, we can evaluate the overall degree of out-of-distribution-ness for an entire system run. Similar to the hybrid SaCFG framework, there is flexibility in choosing whether to discretize states into clusters or to leverage continuous distances in the latent space, allowing for nuanced approaches to modeling and analysis.

\subsection{Debugging ML-based Systems}
\label{sec:debugging}

Once we connect out-of-distribution-ness, unexpected behaviour, and coverage in SFG (or SaCFG), we can consider debugging techniques for traditional programs. For example, we may be able to perform something similar to Spectrum Based Fault Localisation~\cite{Wong:2016aa}: if buggy executions of an LLM-based agent tend to \emph{cover} a specific semantic cluster, while normal executions do not, we may expect that the specific reasoning step relevant to that semantic cluster is the root cause of the problem, and that the related prompt needs to be improved. Alternatively, specific patterns of flow, over multiple clusters, might indicate faulty (multi-step) ``reasoning''.

\subsection{Predicting Execution Results}
\label{sec:prediction}

Kim et al.~\cite{Kim2025aa} initially proposed LLM Inference Graph as a way to predict whether a set of LLM-based agent executions can produce a correct answer: authors have trained a Graph Convolutional Network (GCN) that takes an LIG as an input and predicts whether the final answer is correct or not, achieving precisions of over 0.8. Such predictions are made feasible because LIG, a special case of SFG, encapsulates the entire behaviour of the LLM-based agent system. Since LLMs require very large amount of resources, we argue that such predictions can be very valuable as long as they are reasonable accurate. Further, if accurate predictions can be made using partial SFGs (i.e., SFGs constructed from incomplete executions), we may be able to force early-termination of executions that are not likely to succeed. 

\subsection{Interpretability and Explainability}
\label{sec:interpretability_and_explainability}


The choice of latent spaces, $L_1, \ldots, L_n$, can significantly enhance the interpretability and explainability of ML-based system behavior. Semantic embeddings map specific execution steps, such as an LLM-based agent's response, into latent spaces. We argue that these embeddings can also be tailored to capture domain-specific abstract properties of the responses. For example, in an LLM-based agent providing customized health advice, user inputs could be embedded using a generic language model or into a specially designed latent space. Such a space might distinguish whether the input represents self-reflection, factual information, or a rebuttal to the agent's response. By designing latent spaces to highlight such high-level properties, we can potentially improve the interpretability and explainability of the system's behavior.

\section{Conclusion}
\label{sec:conclusion}

We introduce the concept of semantic flow: the flow of semantic information, represented as vectors in (a) latent space(s), traversed by an ML-based system during execution. This framework applies to systems ranging from individual deep neural networks (DNNs) to complex large language model (LLM) agents, with executions captured and visualized through Semantic Flow Graphs (SFGs).

We discuss properties of this new concept and highlight how semantic flow enables the adaptation of dynamic program analysis techniques for ML-based systems. By using semantic flow to represent executions, we aim to improve the quality, reliability, and understanding of these systems, advancing their analysis and development.

\section*{Acknowledgements}

Shin Yoo, Somin Kim, and Naryeong Kim are supported by the National Research Foundation of Korea (NRF) funded by the Korean Government MSIT (RS-2023-00208998), the Engineering Research Center Program funded by the Korean Government MSIT (RS-2021-NR060080), and the Institute of Information \& Communications Technology Planning \& Evaluation (IITP) grant funded by the Korea government (MSIT) (RS-2022-II220995). Robert Feldt acknowledges support from the Swedish Scientific Council (No. 2020-05272) and from the WASP-funded `BoundMiner' project.

\balance
\bibliographystyle{ACM-Reference-Format}
\bibliography{newref}


\begin{thebibliography}{28}


\ifx \showCODEN    \undefined \def \showCODEN     #1{\unskip}     \fi
\ifx \showDOI      \undefined \def \showDOI       #1{#1}\fi
\ifx \showISBNx    \undefined \def \showISBNx     #1{\unskip}     \fi
\ifx \showISBNxiii \undefined \def \showISBNxiii  #1{\unskip}     \fi
\ifx \showISSN     \undefined \def \showISSN      #1{\unskip}     \fi
\ifx \showLCCN     \undefined \def \showLCCN      #1{\unskip}     \fi
\ifx \shownote     \undefined \def \shownote      #1{#1}          \fi
\ifx \showarticletitle \undefined \def \showarticletitle #1{#1}   \fi
\ifx \showURL      \undefined \def \showURL       {\relax}        \fi
\providecommand\bibfield[2]{#2}
\providecommand\bibinfo[2]{#2}
\providecommand\natexlab[1]{#1}
\providecommand\showeprint[2][]{arXiv:#2}

\bibitem[duo(2024)]%
        {duolingoMax}
 \bibinfo{year}{2024}\natexlab{}.
\newblock \bibinfo{title}{Duolingo Max with GPT-4}.
\newblock
\urldef\tempurl%
\url{https://blog.duolingo.com/duolingo-max/}
\showURL{%
\tempurl}


\bibitem[Gpt(2024)]%
        {Gptrpg2024aa}
 \bibinfo{year}{2024}\natexlab{}.
\newblock \bibinfo{title}{GPTRPG}.
\newblock
\urldef\tempurl%
\url{https://gptrpg.net/en/discover}
\showURL{%
\tempurl}


\bibitem[Brown et~al\mbox{.}(2020)]%
        {Brown2020aa}
\bibfield{author}{\bibinfo{person}{Tom Brown}, \bibinfo{person}{Benjamin Mann},
  \bibinfo{person}{Nick Ryder}, \bibinfo{person}{Melanie Subbiah},
  \bibinfo{person}{Jared~D Kaplan}, \bibinfo{person}{Prafulla Dhariwal},
  \bibinfo{person}{Arvind Neelakantan}, \bibinfo{person}{Pranav Shyam},
  \bibinfo{person}{Girish Sastry}, \bibinfo{person}{Amanda Askell},
  {et~al\mbox{.}}} \bibinfo{year}{2020}\natexlab{}.
\newblock \showarticletitle{Language models are few-shot learners}.
\newblock \bibinfo{journal}{\emph{Advances in neural information processing
  systems}}  \bibinfo{volume}{33} (\bibinfo{year}{2020}),
  \bibinfo{pages}{1877--1901}.
\newblock


\bibitem[Fan et~al\mbox{.}(2023)]%
        {Fan2023yu}
\bibfield{author}{\bibinfo{person}{A. Fan}, \bibinfo{person}{B. Gokkaya},
  \bibinfo{person}{M. Harman}, \bibinfo{person}{M. Lyubarskiy},
  \bibinfo{person}{S. Sengupta}, \bibinfo{person}{S. Yoo}, {and}
  \bibinfo{person}{J.~M. Zhang}.} \bibinfo{year}{2023}\natexlab{}.
\newblock \showarticletitle{Large Language Models for Software Engineering:
  Survey and Open Problems}. In \bibinfo{booktitle}{\emph{Proceedings of the
  45th IEEE/ACM International Conference on Software Engineering: Future of
  Software Engineering}} \emph{(\bibinfo{series}{ICSE-FoSE})}.
  \bibinfo{publisher}{IEEE Computer Society}, \bibinfo{pages}{31--53}.
\newblock
\urldef\tempurl%
\url{https://doi.org/10.1109/ICSE-FoSE59343.2023.00008}
\showDOI{\tempurl}


\bibitem[Farabet et~al\mbox{.}(2013)]%
        {Farabet2013bc}
\bibfield{author}{\bibinfo{person}{Clement Farabet}, \bibinfo{person}{Camille
  Couprie}, \bibinfo{person}{Laurent Najman}, {and} \bibinfo{person}{Yann
  LeCun}.} \bibinfo{year}{2013}\natexlab{}.
\newblock \showarticletitle{Learning hierarchical features for scene labeling}.
\newblock \bibinfo{journal}{\emph{IEEE transactions on pattern analysis and
  machine intelligence}} \bibinfo{volume}{35}, \bibinfo{number}{8}
  (\bibinfo{year}{2013}), \bibinfo{pages}{1915--1929}.
\newblock


\bibitem[{Feng} et~al\mbox{.}(2020)]%
        {Feng2020mz}
\bibfield{author}{\bibinfo{person}{D. {Feng}}, \bibinfo{person}{C.
  {Haase-Sch{\"u}tz}}, \bibinfo{person}{L. {Rosenbaum}}, \bibinfo{person}{H.
  {Hertlein}}, \bibinfo{person}{C. {Gl{\"a}ser}}, \bibinfo{person}{F. {Timm}},
  \bibinfo{person}{W. {Wiesbeck}}, {and} \bibinfo{person}{K. {Dietmayer}}.}
  \bibinfo{year}{2020}\natexlab{}.
\newblock \showarticletitle{Deep Multi-Modal Object Detection and Semantic
  Segmentation for Autonomous Driving: Datasets, Methods, and Challenges}.
\newblock \bibinfo{journal}{\emph{IEEE Transactions on Intelligent
  Transportation Systems}} (\bibinfo{year}{2020}), \bibinfo{pages}{1--20}.
\newblock


\bibitem[{Hinton} et~al\mbox{.}(2012)]%
        {Hinton2012aa}
\bibfield{author}{\bibinfo{person}{G. {Hinton}}, \bibinfo{person}{L. {Deng}},
  \bibinfo{person}{D. {Yu}}, \bibinfo{person}{G.~E. {Dahl}},
  \bibinfo{person}{A. {Mohamed}}, \bibinfo{person}{N. {Jaitly}},
  \bibinfo{person}{A. {Senior}}, \bibinfo{person}{V. {Vanhoucke}},
  \bibinfo{person}{P. {Nguyen}}, \bibinfo{person}{T.~N. {Sainath}}, {and}
  \bibinfo{person}{B. {Kingsbury}}.} \bibinfo{year}{2012}\natexlab{}.
\newblock \showarticletitle{Deep Neural Networks for Acoustic Modeling in
  Speech Recognition: The Shared Views of Four Research Groups}.
\newblock \bibinfo{journal}{\emph{IEEE Signal Processing Magazine}}
  \bibinfo{volume}{29}, \bibinfo{number}{6} (\bibinfo{date}{Nov}
  \bibinfo{year}{2012}), \bibinfo{pages}{82--97}.
\newblock
\showISSN{1053-5888}
\urldef\tempurl%
\url{https://doi.org/10.1109/MSP.2012.2205597}
\showDOI{\tempurl}


\bibitem[Huben et~al\mbox{.}(2024)]%
        {Huben2024aa}
\bibfield{author}{\bibinfo{person}{Robert Huben}, \bibinfo{person}{Hoagy
  Cunningham}, \bibinfo{person}{Logan~Riggs Smith}, \bibinfo{person}{Aidan
  Ewart}, {and} \bibinfo{person}{Lee Sharkey}.}
  \bibinfo{year}{2024}\natexlab{}.
\newblock \showarticletitle{Sparse Autoencoders Find Highly Interpretable
  Features in Language Models}. In \bibinfo{booktitle}{\emph{The Twelfth
  International Conference on Learning Representations}}.
\newblock
\urldef\tempurl%
\url{https://openreview.net/forum?id=F76bwRSLeK}
\showURL{%
\tempurl}


\bibitem[Just et~al\mbox{.}(2014)]%
        {Just:2014aa}
\bibfield{author}{\bibinfo{person}{Ren{\'e} Just}, \bibinfo{person}{Darioush
  Jalali}, {and} \bibinfo{person}{Michael~D. Ernst}.}
  \bibinfo{year}{2014}\natexlab{}.
\newblock \showarticletitle{Defects4J: A Database of Existing Faults to Enable
  Controlled Testing Studies for Java Programs}. In
  \bibinfo{booktitle}{\emph{Proceedings of the 2014 International Symposium on
  Software Testing and Analysis}} (San Jose, CA, USA)
  \emph{(\bibinfo{series}{ISSTA 2014})}. \bibinfo{publisher}{ACM},
  \bibinfo{address}{New York, NY, USA}, \bibinfo{pages}{437--440}.
\newblock
\showISBNx{978-1-4503-2645-2}
\urldef\tempurl%
\url{https://doi.org/10.1145/2610384.2628055}
\showDOI{\tempurl}


\bibitem[Kang et~al\mbox{.}(2024a)]%
        {Kang2024ay}
\bibfield{author}{\bibinfo{person}{Sungmin Kang}, \bibinfo{person}{Gabin An},
  {and} \bibinfo{person}{Shin Yoo}.} \bibinfo{year}{2024}\natexlab{a}.
\newblock \showarticletitle{A Quantitative and Qualitative Evaluation of
  LLM-based Explainable Fault Localization}.
\newblock \bibinfo{journal}{\emph{Proceedings of the ACM on Software
  Engineering}} \bibinfo{volume}{1}, \bibinfo{number}{FSE}
  (\bibinfo{date}{July} \bibinfo{year}{2024}), \bibinfo{pages}{64:1424--1446}.
\newblock


\bibitem[Kang et~al\mbox{.}(2024b)]%
        {Kang2024aa}
\bibfield{author}{\bibinfo{person}{Sungmin Kang}, \bibinfo{person}{Juyeon
  Yoon}, \bibinfo{person}{Nargiz Askarbekkyzy}, {and} \bibinfo{person}{Shin
  Yoo}.} \bibinfo{year}{2024}\natexlab{b}.
\newblock \showarticletitle{Evaluating Diverse Large Language Models for
  Automatic and General Bug Reproduction}.
\newblock \bibinfo{journal}{\emph{IEEE Transactions on Software Engineering}}
  \bibinfo{volume}{50}, \bibinfo{number}{10} (\bibinfo{year}{2024}),
  \bibinfo{pages}{2677--2694}.
\newblock


\bibitem[Kim et~al\mbox{.}(2019)]%
        {Kim2019aa}
\bibfield{author}{\bibinfo{person}{Jinhan Kim}, \bibinfo{person}{Robert Feldt},
  {and} \bibinfo{person}{Shin Yoo}.} \bibinfo{year}{2019}\natexlab{}.
\newblock \showarticletitle{Guiding Deep Learning System Testing using Surprise
  Adequacy}. In \bibinfo{booktitle}{\emph{Proceedings of the 41th International
  Conference on Software Engineering}} \emph{(\bibinfo{series}{ICSE 2019})}.
  \bibinfo{publisher}{IEEE Press}, \bibinfo{pages}{1039--1049}.
\newblock
\urldef\tempurl%
\url{https://doi.org/10.1109/ICSE.2019.00108}
\showDOI{\tempurl}


\bibitem[Kim et~al\mbox{.}(2022)]%
        {Kim2022hg}
\bibfield{author}{\bibinfo{person}{Jinhan Kim}, \bibinfo{person}{Robert Feldt},
  {and} \bibinfo{person}{Shin Yoo}.} \bibinfo{year}{2022}\natexlab{}.
\newblock \showarticletitle{Evaluating Surprise Adequacy for Deep Learning
  System Testing}.
\newblock \bibinfo{journal}{\emph{{ACM} Transactions on Software Engineering
  and Methodology}} \bibinfo{volume}{32}, \bibinfo{number}{2}
  (\bibinfo{date}{June} \bibinfo{year}{2022}), \bibinfo{pages}{1--29}.
\newblock


\bibitem[Kim et~al\mbox{.}(2025)]%
        {Kim2025aa}
\bibfield{author}{\bibinfo{person}{Naryeong Kim}, \bibinfo{person}{Sungmin
  Kang}, \bibinfo{person}{Gabin An}, {and} \bibinfo{person}{Shin Yoo}.}
  \bibinfo{year}{2025}\natexlab{}.
\newblock \showarticletitle{Lachesis: Predicting LLM Inference Accuracy using
  Structural Properties of Reasoning Paths}. In
  \bibinfo{booktitle}{\emph{Proceedings of the 6th International Workshop on
  Deep Learning for Testing and Testing for Deep Learning}}
  \emph{(\bibinfo{series}{DeepTest 2025})}.
\newblock


\bibitem[Krizhevsky(2009)]%
        {Krizhevsky2009kt}
\bibfield{author}{\bibinfo{person}{Alex Krizhevsky}.}
  \bibinfo{year}{2009}\natexlab{}.
\newblock \bibinfo{booktitle}{\emph{Learning Multiple Layers of Features from
  Tiny Images}}.
\newblock \bibinfo{type}{{T}echnical {R}eport}.
  \bibinfo{institution}{University of Toronto}.
\newblock


\bibitem[LeCun et~al\mbox{.}(2015)]%
        {LeCun2015ef}
\bibfield{author}{\bibinfo{person}{Yann LeCun}, \bibinfo{person}{Yoshua
  Bengio}, {and} \bibinfo{person}{Geoffrey Hinton}.}
  \bibinfo{year}{2015}\natexlab{}.
\newblock \showarticletitle{Deep learning}.
\newblock \bibinfo{journal}{\emph{Nature}} \bibinfo{volume}{521},
  \bibinfo{number}{7553} (\bibinfo{year}{2015}), \bibinfo{pages}{436}.
\newblock


\bibitem[Rauber et~al\mbox{.}(2017)]%
        {Rauber2017aa}
\bibfield{author}{\bibinfo{person}{Paulo~E. Rauber}, \bibinfo{person}{Samuel~G.
  Fadel}, \bibinfo{person}{Alexandre~X. Falc{\~a}o}, {and}
  \bibinfo{person}{Alexandru~C. Telea}.} \bibinfo{year}{2017}\natexlab{}.
\newblock \showarticletitle{Visualizing the Hidden Activity of Artificial
  Neural Networks}.
\newblock \bibinfo{journal}{\emph{IEEE Transactions on Visualization and
  Computer Graphics}} \bibinfo{volume}{23}, \bibinfo{number}{1}
  (\bibinfo{year}{2017}), \bibinfo{pages}{101--110}.
\newblock
\urldef\tempurl%
\url{https://doi.org/10.1109/TVCG.2016.2598838}
\showDOI{\tempurl}


\bibitem[Reimers and Gurevych(2019)]%
        {Reimers2019aa}
\bibfield{author}{\bibinfo{person}{Nils Reimers} {and} \bibinfo{person}{Iryna
  Gurevych}.} \bibinfo{year}{2019}\natexlab{}.
\newblock \showarticletitle{Sentence-{BERT}: Sentence Embeddings using
  {S}iamese {BERT}-Networks}. In \bibinfo{booktitle}{\emph{Proceedings of the
  2019 Conference on Empirical Methods in Natural Language Processing and the
  9th International Joint Conference on Natural Language Processing
  (EMNLP-IJCNLP)}}, \bibfield{editor}{\bibinfo{person}{Kentaro Inui},
  \bibinfo{person}{Jing Jiang}, \bibinfo{person}{Vincent Ng}, {and}
  \bibinfo{person}{Xiaojun Wan}} (Eds.). \bibinfo{publisher}{Association for
  Computational Linguistics}, \bibinfo{address}{Hong Kong, China},
  \bibinfo{pages}{3982--3992}.
\newblock
\urldef\tempurl%
\url{https://doi.org/10.18653/v1/D19-1410}
\showDOI{\tempurl}


\bibitem[Sharifi et~al\mbox{.}(2023)]%
        {Sharifi2023aa}
\bibfield{author}{\bibinfo{person}{Sepehr Sharifi}, \bibinfo{person}{Donghwan
  Shin}, \bibinfo{person}{Lionel~C. Briand}, {and} \bibinfo{person}{Nathan
  Aschbacher}.} \bibinfo{year}{2023}\natexlab{}.
\newblock \showarticletitle{Identifying the Hazard Boundary of ML-Enabled
  Autonomous Systems Using Cooperative Coevolutionary Search}.
\newblock \bibinfo{journal}{\emph{IEEE Transactions on Software Engineering}}
  \bibinfo{volume}{49}, \bibinfo{number}{12} (\bibinfo{year}{2023}),
  \bibinfo{pages}{5120--5138}.
\newblock
\urldef\tempurl%
\url{https://doi.org/10.1109/TSE.2023.3327575}
\showDOI{\tempurl}


\bibitem[Vaswani et~al\mbox{.}(2017)]%
        {Vaswani2017aa}
\bibfield{author}{\bibinfo{person}{Ashish Vaswani}, \bibinfo{person}{Noam
  Shazeer}, \bibinfo{person}{Niki Parmar}, \bibinfo{person}{Jakob Uszkoreit},
  \bibinfo{person}{Llion Jones}, \bibinfo{person}{Aidan~N Gomez},
  \bibinfo{person}{\L ukasz Kaiser}, {and} \bibinfo{person}{Illia Polosukhin}.}
  \bibinfo{year}{2017}\natexlab{}.
\newblock \showarticletitle{Attention is All you Need}. In
  \bibinfo{booktitle}{\emph{Advances in Neural Information Processing
  Systems}}, \bibfield{editor}{\bibinfo{person}{I.~Guyon},
  \bibinfo{person}{U.~Von Luxburg}, \bibinfo{person}{S.~Bengio},
  \bibinfo{person}{H.~Wallach}, \bibinfo{person}{R.~Fergus},
  \bibinfo{person}{S.~Vishwanathan}, {and} \bibinfo{person}{R.~Garnett}}
  (Eds.), Vol.~\bibinfo{volume}{30}. \bibinfo{publisher}{Curran Associates,
  Inc.}
\newblock


\bibitem[Wang et~al\mbox{.}(2023)]%
        {Wang2023aa}
\bibfield{author}{\bibinfo{person}{Xuezhi Wang}, \bibinfo{person}{Jason Wei},
  \bibinfo{person}{Dale Schuurmans}, \bibinfo{person}{Quoc Le},
  \bibinfo{person}{Ed Chi}, \bibinfo{person}{Sharan Narang},
  \bibinfo{person}{Aakanksha Chowdhery}, {and} \bibinfo{person}{Denny Zhou}.}
  \bibinfo{year}{2023}\natexlab{}.
\newblock \showarticletitle{Self-Consistency Improves Chain of Thought
  Reasoning in Language Models}.
\newblock \bibinfo{journal}{\emph{CoRR}}  \bibinfo{volume}{abs/2203.11171}
  (\bibinfo{year}{2023}).
\newblock


\bibitem[Wong et~al\mbox{.}(2016)]%
        {Wong:2016aa}
\bibfield{author}{\bibinfo{person}{W.~E. Wong}, \bibinfo{person}{Ruizhi Gao},
  \bibinfo{person}{Yihao Li}, \bibinfo{person}{Rui Abreu}, {and}
  \bibinfo{person}{Franz Wotawa}.} \bibinfo{year}{2016}\natexlab{}.
\newblock \showarticletitle{A Survey on Software Fault Localization}.
\newblock \bibinfo{journal}{\emph{IEEE Transactions on Software Engineering}}
  \bibinfo{volume}{42}, \bibinfo{number}{8} (\bibinfo{date}{August}
  \bibinfo{year}{2016}), \bibinfo{pages}{707}.
\newblock


\bibitem[Yan et~al\mbox{.}(2023)]%
        {Yan2023aa}
\bibfield{author}{\bibinfo{person}{Ming Yan}, \bibinfo{person}{Ruihao Li},
  \bibinfo{person}{Hao Zhang}, \bibinfo{person}{Hao Wang},
  \bibinfo{person}{Zhilan Yang}, {and} \bibinfo{person}{Ji Yan}.}
  \bibinfo{year}{2023}\natexlab{}.
\newblock \showarticletitle{Larp: Language-agent role play for open-world
  games}.
\newblock \bibinfo{journal}{\emph{arXiv preprint arXiv:2312.17653}}
  (\bibinfo{year}{2023}).
\newblock


\bibitem[Yang et~al\mbox{.}(2024)]%
        {Yang2024ab}
\bibfield{author}{\bibinfo{person}{John Yang}, \bibinfo{person}{Carlos~E.
  Jimenez}, \bibinfo{person}{Alexander Wettig}, \bibinfo{person}{Kilian
  Lieret}, \bibinfo{person}{Shunyu Yao}, \bibinfo{person}{Karthik Narasimhan},
  {and} \bibinfo{person}{Ofir Press}.} \bibinfo{year}{2024}\natexlab{}.
\newblock \bibinfo{title}{SWE-agent: Agent-Computer Interfaces Enable Automated
  Software Engineering}.
\newblock
\showeprint[arxiv]{arXiv.2405.15793}~[cs.SE]
\urldef\tempurl%
\url{https://arxiv.org/abs/2405.15793}
\showURL{%
\tempurl}


\bibitem[Yao et~al\mbox{.}(2023)]%
        {Yao2022qf}
\bibfield{author}{\bibinfo{person}{Shunyu Yao}, \bibinfo{person}{Jeffrey Zhao},
  \bibinfo{person}{Dian Yu}, \bibinfo{person}{Nan Du}, \bibinfo{person}{Izhak
  Shafran}, \bibinfo{person}{Karthik Narasimhan}, {and} \bibinfo{person}{Yuan
  Cao}.} \bibinfo{year}{2023}\natexlab{}.
\newblock \showarticletitle{ReAct: Synergizing Reasoning and Acting in Language
  Models}. In \bibinfo{booktitle}{\emph{Proceedings of the International
  Conference on Learning Representation}} \emph{(\bibinfo{series}{ICLR 2023})}.
\newblock


\bibitem[Yosinski et~al\mbox{.}(2015)]%
        {Yosinski2015oh}
\bibfield{author}{\bibinfo{person}{Jason Yosinski}, \bibinfo{person}{Jeff
  Clune}, \bibinfo{person}{Anh~Mai Nguyen}, \bibinfo{person}{Thomas~J. Fuchs},
  {and} \bibinfo{person}{Hod Lipson}.} \bibinfo{year}{2015}\natexlab{}.
\newblock \showarticletitle{Understanding Neural Networks Through Deep
  Visualization}.
\newblock \bibinfo{journal}{\emph{CoRR}}  \bibinfo{volume}{abs/1506.06579}
  (\bibinfo{year}{2015}).
\newblock
\showeprint[arXiv]{1506.06579}
\urldef\tempurl%
\url{http://arxiv.org/abs/1506.06579}
\showURL{%
\tempurl}


\bibitem[Yun et~al\mbox{.}(2021)]%
        {Yun2021aa}
\bibfield{author}{\bibinfo{person}{Zeyu Yun}, \bibinfo{person}{Yubei Chen},
  \bibinfo{person}{Bruno~A Olshausen}, {and} \bibinfo{person}{Yann LeCun}.}
  \bibinfo{year}{2021}\natexlab{}.
\newblock \showarticletitle{Transformer visualization via dictionary learning:
  contextualized embedding as a linear superposition of transformer factors}.
\newblock \bibinfo{journal}{\emph{arXiv preprint arXiv:2103.15949}}
  (\bibinfo{year}{2021}).
\newblock


\bibitem[Zhang et~al\mbox{.}(2018)]%
        {Zhang2018ab}
\bibfield{author}{\bibinfo{person}{Quanshi Zhang}, \bibinfo{person}{Ying~Nian
  Wu}, {and} \bibinfo{person}{Song-Chun Zhu}.} \bibinfo{year}{2018}\natexlab{}.
\newblock \showarticletitle{Interpretable Convolutional Neural Networks}. In
  \bibinfo{booktitle}{\emph{Proceedings of the IEEE Conference on Computer
  Vision and Pattern Recognition (CVPR)}}.
\newblock


\end{thebibliography}

\end{document}